\documentclass[11pt]{article}
\usepackage{amssymb}

\begin{document}

\author{A.A. Suzko $^{a}$\footnote{email: suzko@jinr.ru}, I. Tralle $^{b}$\footnote
{email; tralle@univ.rzeszow.pl}\\
\em {$^a$ Joint Institute for Nuclear Research, 141980 Dubna, Russia}\\
\em {$^{b}$ Institute of Physics,Mathematics and Natural Sciences Department}\\
\em {University of Rzesz\'ow 35-310 Rzesz\'ow, Poland}}

 \newpage
 \begin{center}
{\Large{\bf
 Reconstruction of Quantum Well Potentials via  the Intertwining Operator
Technique}}
 \end{center}

\vskip0.4cm
 \begin{center}
{\large \bf A. A. Suzko}$^{a}$\footnote{email: suzko@jinr.ru},
{\large \bf I. Tralle }$^{b}$\footnote {email;
tralle@univ.rzeszow.pl}

{$^a$ Joint Institute for Nuclear Research, 141980 Dubna, Russia}\\
{$^{b}$ Institute of Physics,Mathematics and Natural Sciences Department}\\
{University of Rzesz\'ow 35-310 Rzesz\'ow, Poland}
\end{center}

\begin{abstract}

    One of the most important issues of {\it quantum engineering} is the construction of
    low-dimensional structures possessing  desirable properties. For example, in different
    areas of possible applications of the structures containing quantum wells (QW),
    there is need to have QW energy spectrum, which is predetermined.
    Then the following question arises:
    can one reconstruct the shape of QW which supports this spectrum?
    We outline the possible strategy of the QW potential shape reconstruction, if
    the spectrum of QW is given in advance. The proposed approach is based on the
     combination of different techniques such as {\it Inverse Scattering Problem Method},
     {\it Darboux} and {\it Liouville} transformation. It enables to take into
     account the space-variable dependent effective mass of charge carriers and allows
     the kinetic energy operator to be of  non-Hermitian as well as Hermitian form.

The proposed technique allows  to construct phase-equivalent
potentials, to add the new bounded states to (or remove some of
them from) the spectrum supported by  an initial potential and
provides a systematic procedure for generating new exactly
solvable models.
\\

\end{abstract}

\section{Introduction}

        In recent years we have been  witnesses of the rapid progress in
        {\it nanoelectronics}
        which is already on the way to continuing the outstanding successes of
        microelectronics. This became possible among others, due to the development
         of technologies and techniques, such as {\it Molecular Beam Epitaxy (MBE)}
for instance, which enables to deposit thin layers of different
materials one on top of the other,
         with almost atomic precision. The last one, in its turn, is able to produce
         a variety  of  {\it low-dimensional structures}, ranging from a heterojunction
          formed at a single interface, through quantum wells (QW) to superlattices.
          It would not be an overstatement to say that a new paradigm of electronics
          emerged, for which even the name has been  already coined,
           {\it Quantum Technology} or {\it Quantum Engineering}. It seems, however,
           that quantum engineering in its present stage, in spite of all its successes
 and maturity, is still {\it passive} in the sense that it makes use of, figuratively
 speaking, less 'degrees of freedom' than it possibly could. It means that the 'palette'
 of QW-potential shapes is still limited to a few most popular ones: rectangular,
  parabolic or semi-parabolic and this circumstance obviously restricts the possibility
  to choose and control the energy spectrum of QW produced by means of MBE. Meantime, in
  different areas of possible applications of the low-dimensional structures mentioned
  above, there is often need to have some specific kind of spectrum known beforehand and
  a   question arises: how to produce the QW with a predetermined spectrum?
   At least from the theorist's point of view, this question can be
   reformulated as follows: suppose the spectrum of QW is known; that is, one knows
   the number of quantum levels and the distances between them (remember, the spectrum is
   not necessarily  equidistant as in case of a parabolic well or similar to the spectrum of
   a square well); can one reconstruct the QW-potential which supports this spectrum?
    An affirmative answer to the question would make quantum engineering more flexible
    and {\it active}, providing the opportunity to develop multitude of novel quantum devices.

    The aim of this paper is to develop an approach to the QW-potential reconstruction
    which combines different techniques, such as {\it Inverse Scattering Problem Method},
    {\it Darboux} (or intertwining operator technique) and {\it Liouville} transformations
    in reference to the generalized (it means, with space-variable dependent mass)
    Schr\"odinger equation.

\section{Necessary preliminaries}

    As we already mentioned, one often needs to know, how to reconstruct the potential
provided that the spectrum supported by this potential is known
beforehand. One can look at the problem also in this way: suppose
we have a set of "scattering data", whatever it means; we do not
define them precisely right now. Can we find the potential which
produces this set of data ?
 Affirmative answer to this question was obtained
for the first time by  mathematicians I.M. Gel'fand, B.M. Levitan
[1,2] and V.A. Marchenko [3] (see also Refs. therein). They
elaborated the method of potential reconstruction by means of
spectral or scattering data which is now known under the name of
{\it Inverse Scattering Problem Method (ISP)}. We shall term their
method or approach also as a GLM-method (approach). $\;$ In a
particular case of $QW$, the question is reduced to the following:
suppose we have a number of quantum levels with the given
distances between them; what the potential shape (or which the
form of) $QW$ should be in this case; or in other words, can we
find the potential of $QW$,  if the spectrum and the depth of $QW$
are given? In Ref.[4] the provisional positive answer to this
question was obtained by means of the ISP-method. The reader also
can find in Ref. [4] the details concerning the motivation for
such inquiry. Here, for the sake of consistency and reader's
convenience, we outline  the main ideas of GLM-approach only very
briefly.

Consider the eigenvalue problem of the Schr\"{o}dinger equation
with the potential $V(x)$ in one space dimension
\begin{equation}
\left(-\frac{d^{2}}{d x^{2}} + V(x)\right)\phi(x,k) =
k^{2}\phi(x,k),
\end{equation}
(here we assume $\hbar^2/2m=1$). The GLM-method may be viewed as a
dispersion theory for the Schr\"{o}dinger wave function of (1).
>From the solution to (1) with $k$ complex, one can define Jost
functions $f_{1}^{\pm}$ and $f_{2}^{\pm}$ which are analytic in
the upper-half $k$-plane with the following asymptotic behaviour
 \begin{displaymath}
 f_{1}^{\pm}(x,k) \sim \exp(\pm ikx),{\rm as}\; x\rightarrow +\infty,
 \end{displaymath}
 \begin{displaymath}
 f_{2}^{\pm}(x,k) \sim \exp(\mp ikx),{\rm as}\; x\rightarrow -\infty,
 \end{displaymath}
and construct a meromorphic function $\Phi(x,k)$ as
\begin{displaymath}
\Phi(x,k) = a^{-1}(k)f_{2}^{+}(x,k)\exp(ik x), {\rm
Im}k > 0;\end{displaymath}
\begin{displaymath}
 =f_{1}^{-}(x,k^{*})\exp(ik x), {\rm Im}k < 0, \end{displaymath}
where $a^{-1}(k)$ is the conventional transmission
coefficient.

Thus, $\Phi(x,k)$ is completely determined by its singularity
structure which consists of a cut along the real $k$-axis and some
number, say $N$ of bound-state poles on the positive imaginary
axis. The spectral weight of the cut is essentially a
scattering-state wave function multiplied by the reflection
coefficient, both evaluated at real $k$. Similarly, the pole
residues are essentially constants times bound-state wave
functions. Upon Fourier transformation the dispersion relation for
$\Phi$ becomes the Marchenko integral equation [3] (see also [5]
for details) which determines the wave functions. The inverse
problem of the potential reconstruction by means of scattering
data can be reduced, generally speaking, to solving an  integral
equation
\begin{displaymath}
K(x,x') + Q(x,x') + \int_{x}^{\infty}K(x,x^{''})Q(x^{''},x')dx^{''}=0
\end{displaymath}
and thereupon, to a simple differentiation of its kernel
\begin{displaymath} V(x)=-2\frac{d}{dx}K(x,x) \end{displaymath}
which is given entirely in terms of the reflection coefficient and
$2N$ bound-state parameters. Here \begin{displaymath} Q(x,x')=
\frac{1}{2\pi}\int_{-\infty}^{\infty}[1-S(k)]\exp (ik(x+x'))dk +
\sum_{n}^{N}M_n^{2} \exp(-\kappa_n(x+x')),\end{displaymath}
 where
$S(k)$, $E_n$ and $M^{2}_n$ are  scattering data: $S(k)$ is a
scattering matrix, $E_n=(i\kappa_n)^2$ are bound-state energies
and  $M^{2}_n$ are its normalization constants.

If the reflection coefficient can be represented by rational
functions of $k$, the Marchenko equation can be solved exactly by
algebraic technique. The most simple case corresponds to a
reflection coefficient vanishing for all real $k$ ($S(k)=1$); then
the expression for $Q(x,x')$ contains only the sum over the bound
states,  and the integral equation reduces to a system of $N$
linear algebraic equations with respect to the kernel $K(x,x')$.
The potential $V(x)$ also can be reconstructed by means of $2N$
parameters. The half of the parameters  makes $N$ bound-state
energies $E_n$, $n=1,2,...N$, while the others are the normalizing
constants $M^{2}_n$ which in {\it Nuclear Physics} are assumed  to
be obtained from scattering data as $M^{2}_n=iRes
S(k)|k=i\kappa_{n}$. It corresponds to the transparent and
symmetric potentials. These $2N$ numbers supply a complete set of
the parameters for solving the inverse problem to exist and be
unique. Obviously, in case of $QW$s the second half of the
scattering data (the normalizing constants) is inaccessible. It
turns out however, that the way out actually exists. The
possibility to reconstruct the infinitely deep and symmetric
potentials solely by means of bounded states was established for
the first time by B.M Levitan and M.G. Gasymov who proved the {\it
The Theorem about Two-spectrum } [6,2]. The meaning of this
theorem in brief is the following: to reconstruct symmetric
potentials $V(x)=V(-x)$, it suffices to know only the positions of
energy levels without knowing the normalizing constants. Later on,
this fact was re-discovered by H.B.Thacker, C.Quigg and J.L.Rosner
[7] who modelled the confining potentials binding massive quarks
and antiquarks in meson systems. They were able even to fit
approximately the lower parts of infinitely deep wells (parabolic,
harmonic oscillator, linear and rectangular wells) by means of a
limited number of levels whose positions are known, using some
additional technical trick which is termed  a {\it stabilizing
parameter}.
 This observation was used in Ref.[4] in order to
reconstruct the shape of QW provided that its energy spectrum and
depth are known in advance. The problem however is, that in the
approach developed there, the electron effective mass was supposed
to be constant and independent on the space variable, while in
reality it is space-dependent. It is because in practice the
quantum wells of different shapes are produced by means of
MBE technique, when the semiconductor layers
are grown subsequently one by one and these layers are
characterized by different electron effective masses. The
subsequent semiconductor layers stuck together make the
quantum well and as a result, one can consider such structure as
having effective mass which depends on the space variable. Proceed
now to the thorough treatment of this problem.

\section{Reconstruction of quantum well potentials using intertwining operator technique}
\subsection{ISP and reconstruction of quantum well potentials}

The discussion presented above suggests the following strategy of
$QW$ potential reconstruction. At first, one has to search for the
solution of our problem  among the potentials of a special class
of reflectionless and symmetric potential $V(x)=V(-x)$, using the
{\it ISP}-\rm method as it was done in [4]. Then  one can amend
the potential reconstructed in this way, using the intertwining
operator technique  and taking into account the dependence of the
electron effective mass on the space variable.  Now we are able to
describe this recipe in more details. Suppose that one-dimensional
potential $V(x)$ can be represented by the function
$V_{N}(x,m^{*},E_{0})$ which  obeys the
following conditions:\\
(i) $V_{N}$ supports precisely $N$ bounded states of the quantum
system with the effective mass $m^{*}$. The bound-state energies
coincide with the energies
$\epsilon_{1},\epsilon_{2},...\epsilon_{N}$ of the levels within the $QW$ ;\\
(ii) \begin{math} \lim_{x\rightarrow \infty} V_{N} = E_{0}
\end{math}. The last value can be considered as the depth of $QW$.

Arranging the binding energies $k^{2} = E_{0} - \epsilon_{n}$ in
descending order so that $k_{1} > k_{2} >...> k_{N}$, and
$\epsilon_{1} =E_{0} - k^{2}_{1}$ refers to the ground-state
energy, one can use for  $QW$ potential reconstruction the
technique developed by Schonefeld \em et al \rm [8] for studying
the convergence of the reflectionless approximation to the
confining potentials. Omitting the intermediate calculations, we
give here only the final results:
\begin{equation}
V_{N}(x,E_{0})=E_{0}-2\frac{d^{2}}{dx^{2}}\ln D(x),
\end{equation}
where
\begin{displaymath}
D(x)=\sum_{S}\exp(-2x\sum_{p\in S}k_{p})\prod (S,\tilde{S})
\end{displaymath}
\begin{displaymath}
\prod (S,\tilde{S}) = \prod_{m\in S,n\in \tilde{S}} \frac{k_{m} +
k_{n}}{k_{m}-k_{n}}.
\end{displaymath}
Here the sum ranges over all subsets $S$ of
$\left\{1,2,...,N\right\}$ including the null set and the full
set, while $\tilde{S}$ denotes the complement of the set $S$.

  Now we consider a {\it generalized} Schr\"odinger equation, that is with the
  potential $V_N(x,E_0)$ and the position-dependent effective mass. Obviously, the new
  spectrum obtained in this way  might differ from that which was used in
ISP-method in order to reconstruct the potential $V_N(x,E_0)$, but
we can suppose that the changes of a spectrum are not dramatic,
because in practice the space variable dependence of the effective
mass is weak. Further on we shall show that it is possible to
amend this potential and obtain "improved" one $\widetilde{V}(x)
$, in order to have the spectrum  needed.

\subsection{First-order Darboux transformation and supersymmetry}

The problem of the space-variable dependent effective mass
attracts now persistent attention because it is not obvious
whether the effective
 mass approximation is applicable to heterostructures,
 or not (see [9,10] and the Refs. therein). Before we
 start our discussion, let us  make some general remarks. Remember, the QW of the shape
  other than rectangular, is produced by stacking up a number (some times even
  a considerable number) of layers of different materials, each of which is characterized
   by its own effective mass. This stack of layers can be considered as some special case
    of heterostructure. Then if one tries to solve the Sturm-Liouville problem for
   a corresponding  Schr\"odinger equation (we refer to this problem also
   as a
    {\it direct} one), treating the heterostructure as a whole and applying
    the effective-mass theory, one encounters some difficulties, whose nature is the
    following. First, the total electron wave function is a product of the slowly
    varying {\it envelope} function and the Bloch function of the local extremum in the
    host's band structure. The Bloch functions in the two materials on either side of
    a heterojunction must be similar for the effective-mass approximation to be valid.
     An obvious condition is that they must belong to the same point in Brillouin zone,
     and this can fail for some materials. The second point concerns the matching of the
     envelope functions at the interface. Consider a junction at $x=0$ between two regions of
     materials, say $A$ and $B$. The
Schr\"odinger equations for the envelope function in the two subsequent regions
(we consider only one-dimensional model), are
\begin {displaymath} \left ( - \frac{1}{m^*_A}\frac{d^2}{dx^2} + E^A_c\right) \phi(x)
={\cal E}\chi(x)),\end{displaymath}
\begin {displaymath} \left ( - \frac{1}{m^*_B}\frac{d^2}{dx^2} + E^B_c\right) \phi(x)
={\cal E}\chi(x)),\end{displaymath} where $m^*_A$ and $m^*_B$ are
the electron effective masses for the materials $A$ and $B$,
respectively, $\hbar^2/2=1$ and the difference in the bottoms of
the conduction bands is $\Delta E_c=E^B_c-E^A_c$. If the materials
were the same, one can match the value and the derivative of the
wave function at the interface by means of usual conditions
\begin{displaymath} \phi^A(0_{-})=\phi^B(0_{+}),~~~
\frac{d\phi^A(x)}{dx}|_{x=0_{-}}=\frac{d\phi^B(x)}{dx}|_{x=0_{+}},\end{displaymath}
where $0_{-}$ means the side of the interface in material $A$ and
so on. This simple condition is not correct for the
heterostructure where the two effective masses are different,
because it does not conserve current. A correct set of matching
conditions is
\begin{displaymath} \phi^A(0_{-})=\phi^B(0_{+}),~~
\frac{1}{m^*_A}\frac{d\phi^A(x)}{dx}|_{x=0_{-}}=
\frac{1}{m^*_B}\frac{d\phi^B(x)}{dx}|_{x=0_{+}}.\end{displaymath}
The condition for matching the derivative now includes the
effective mass. A more mathematical argument is that the matching
condition which does not include effective masses assumes that the
Schr\"odinger equation takes the form
\begin{displaymath}
 -\frac{1}{m^*(x)}\frac{d^2\phi}{dx^2}+V(x)\phi(x)={\cal E}\phi(x).
 \end{displaymath}
 This is not Hermitian (or Sturm-Liouville form) if $m^*(x)$ varies, and many of the
 crucial properties of the wave functions might disappear as a result. It seems that
  a consensus concerning the form of Hermitian kinetic energy operator in case of the
  position-dependent effective mass is already achieved among the specialists
  (see Refs. [9,10] and Sec. 4 below). In this section however, we are dealing with
  the intertwining relation technique, which is universal and apply this technique
  to construct a chain
  of exactly solvable Hamiltonians whose kinetic energy operators are not Hermitian, if
  $m^*(x)$ is space dependent.
  In the next section (Sec. 4), we also develop  an approach which gives the
  possibility to
  treat the case of Hermitian kinetic energy operator on the same footing.

Let us start with the equation:
  \begin{equation} \label{H-eq}
  {\cal H}\phi(x)={\cal E}\phi(x),\qquad {\cal H}=-\frac{1}{m^*(x)}\frac{d^2}{dx^2} +
V(x),\end{equation} where $m^*(x)$ is a position dependent
"effective mass" and $V(x)$ is supposed to be equal to
$V_{N}(x,E_0)$. This equation is reduced to the generalized
Schr\"odinger equation of the form:
\begin{displaymath}
 {\cal H}_0\phi_0(x)={\cal E}m^*(x)\phi_0(x),\qquad
{\cal H}_0=-d^2/dx^2 + v(x),\end{displaymath} where
$v(x)=V(x)m^*(x)$. In fact, it is the Schr\"odinger equation with
linearly energy-dependent potentials. The Darboux transformations
for the Schr\"odinger  equations with variable values of energy
and angular momentum  were suggested in [11] and in a more general
form in [12]. Then in Refs. [13,14] algebraic transformations have
been elaborated for a Sturm-Liouville problem for studying
phase-equivalent linearly energy-dependent potentials and for
constructing exactly solvable three-body models with two-central
potentials. On the other hand, the intertwining operator method
provides the universal approach to creating new exactly solvable
models and can be applied to the operators of a very general form
( see for example [15-17]). In this paper, we apply the
intertwining operator technique to the equation (3) with a
position-dependent mass in order to construct the potential which
supports the desirable spectrum.

Suppose that the solution of the eigenvalue problem to the
equation (\ref{H-eq}) with the given potential $V(x)$ and position
dependent $m^*(x)$ are known and we would like to solve a similar
problem for another Hamiltonian $\widetilde{\cal H}$ containing a
new potential ${\widetilde V}(x)$ and  the spectrum which probably
differs from the spectrum of the Hamiltonian (\ref{H-eq}) by a
single quantum state:
\begin{equation}\label{H1-eq}
{\widetilde{\cal H}}{\widetilde{\phi}}(x)={\cal E}{\widetilde{\phi}}(x),\qquad
 {\widetilde{\cal H}}=-\frac{1}{m^*(x)}\frac{d^2}{dx^2} + {\widetilde V}(x).
 \end{equation}

 We start with  standard intertwining relations (see, for instance [15,16]):
\begin{equation}
 {\cal L}{\cal H}={\widetilde{\cal H}}{\cal L},\end{equation}
 \begin{equation} {\widetilde{\phi}}(x)={\cal L}\phi(x),\end{equation}
 where the operator ${\cal L}$ intertwines the Hamiltonians ${\cal H}$ and
 ${\widetilde{\cal H}}$. We search for the
intertwining operator $\cal L$ in a general form
\begin{equation}
\label{L} {\cal L}=B(x)d/dx + A(x),\end{equation}
 where
$A(x)$ and $B(x)$ are to be determined. Once the operator ${\cal
L}$ is known, the solutions ${\widetilde{\phi}}$ can be obtained
from (6) by applying ${\cal L}$ to the known solutions $\phi$. To
find the explicit form of ${\cal L}$, we use the equations
(\ref{H-eq}),(4) and the intertwining relations (5),(6):
 \begin{displaymath} \label{B1}
 \Bigl[-\frac{1}{m^*(x)}
\frac{d^2}{dx^2} + {\widetilde{V}}(x)\Bigr]{\cal L}\phi(x)={\cal
L}\Bigl[-\frac{1}{m^*(x)}\frac{d^2}{dx^2} +
V(x)\Bigr]\phi(x).\end{displaymath} After some algebra we arrive
at:
\begin{displaymath} -\frac{1}{m^*}(A^{''}\phi + 2A^{'}\phi^{'} + A\phi^{''})
 - \frac{1}{m^*}(B^{''}\phi^{'} +
2B^{'}\phi^{''} + B\phi^{'''})+ {\widetilde V}(A\phi +B\phi^{'})=\end{displaymath}
\begin{displaymath}A\Bigl(-\frac{1}{m^*}\phi^{''}+ V\phi\Bigr) +
B\Bigl(-\frac{1}{m^*}\phi^{'''}-
B\Bigl(\frac{1}{m^*}\Bigr)^{'}\phi^{''} + V^{'}\phi +
V\phi^{'}\Bigr)\end{displaymath} and finally, to the next system
of equations:
\begin{equation} \frac{1}{m^*}A + 2\frac{1}{m^*}B^{'} = B\Bigl(\frac{1}{m^*}\Bigr)^{'}
+ A\frac{1}{m^*},\end{equation}
\begin{equation} \frac{1}{m^*}2A^{'} + \frac{1}{m^*}B^{''} - {\widetilde V}B = -BV,
\end{equation}
\begin{equation}- \frac{1}{m^*}A^{''} + {\widetilde V}A =
AV + BV^{'}.\end{equation} From (8) it immediately follows  that
\begin{equation} \label{B}
2B^{'}/B = -m^{*'}/m^*, B= C/\sqrt{m^*} ,\end{equation} where $C$
is an arbitrary constant. From (9), (10) one gets
\begin{equation}\label{Vtilde}
 {\widetilde
V}= V + \frac{1}{m^*}\frac{B^{''}}{B} +
\frac{1}{m^*}\frac{2A^{'}}{B}
\end{equation}
and \begin{displaymath} -\frac{1}{m^*}A{''} +
(\frac{1}{m^*}2A^{'}+ B^{''})B^{-1}A = BV^{'}.\end{displaymath}
 In order to integrate the last equation, let us introduce a new auxiliary
function $K(x)$ defined as $A(x)=B(x)K(x)$. Then we arrive at a
nonlinear differential equation
\begin{displaymath} \left(-K^{''}+ 2K^{'}K - V^{'}m^*\right) + \frac{2B^{'}}
{B}\left(K^2 - K^{'}\right)=0. \end{displaymath}
Taking into
account the relation $V=v/m^*$ and the first of the relations
(11), the last equation can be easily transformed into another
one, in a single unknown $K$ only:
\begin{displaymath} \left( -K^{''} + 2K^{'}K -v^{'}\right)- \frac{m^{*'}}{m^*}\left(
-K^{'} + K^2 - v\right)=0.\end{displaymath}
This one can be rewritten as
\begin{displaymath} \frac{d}{dx}\left( \frac{1}{m^*}\left(-K^{'}+ K^2 - v\right)\right)=0,
\end{displaymath} which means that
\begin{displaymath} (1/m^*)\left( -K^{'}+ K^2 -
v\right)=\nu ,\end{displaymath}
 where $\nu$ is an integration constant.
The last equation is analogous to \it Riccati \rm equation.
Introducing a new function ${\cal U}(x)$ as $K=-{\cal U^{'}}{\cal
U}^{-1}$ and changing $\nu=-\lambda$,  one arrives at the equation
\begin{equation}\label{Eq-U}
-(1/m^*(x)){\cal U^{''}}(x)+ V(x){\cal U}(x)=\lambda{\cal
U}(x).\end{equation} Here ${\cal U}(x)$ is supposed to be
invertible at all $x$. The last equation then is nothing else but
the initial equation (\ref{H-eq}) which is supposed to be solved
and ${\cal E}=\lambda $ is a point of spectrum of ${\cal H}$.
Therefore, we assume that the solutions of (13) are known for the
given values of $\lambda$. Having found the explicit form of $B$
(see
 (11)), using the formula for $K$ mentioned above,  from the
relation $A=BK$ one gets $A(x)=-C\left (\ln {\cal U}(x)\right)^{'}
\sqrt{1/m^*(x)}$. Once ${\cal U}$ is known, the transformation
operator ${\cal L}$, the new potential $\widetilde{V}(x)$ and the
corresponding solutions of the transformed equation (4) are
defined up to an arbitrary constant $C$. Without loss of
generality, we can put it safely equal to unity. After this we
have
\begin{equation}\label{BAK}
B(x)=1/\sqrt{m^*(x)}, ~~ A(x)=K/\sqrt{m^*(x)}, ~~K=- \left (\ln
{\cal U}(x)\right)^{'}.\end{equation} To make further
transformations, let us  calculate
$B''/B=\sqrt{m^*}(1/\sqrt{m^*})''$. Using this and (\ref{BAK}) in
(7), (12) and (6) we  construct the intertwining operator ${\cal
L}$, the transformed potential $\widetilde{V}(x)$ and the
solutions $\widetilde{\phi}$  in the form:
\begin{equation}
\label{L1} {\cal L}=\frac{1}{\sqrt{m^*}}\left(\frac{d}{dx}+
K\right)= \frac{1}{\sqrt {m^*}} \left(\frac{d}{dx} - (\ln{\cal
U}){'}\right),
\end{equation}
\begin{eqnarray}\label{trV}
& &{\widetilde V}= V+ \frac{1}{\sqrt
{m^*}}\left[\frac{d^2}{dx^2}\frac{1}{\sqrt {m^*}}+
2\frac{d}{dx}\left( \frac{1}{\sqrt {m^*}}K \right)\right]=\\
 V+ \frac{1}{\sqrt {m^*}}\left[\frac{d^2}{dx^2}\frac{1}{\sqrt
{m^*}}- 2\frac{d}{dx}\left( \frac{1}{\sqrt {m^*}}(\ln{\cal U}){'}
\right)\right],\nonumber\end{eqnarray}
\begin{equation}\label{w-phi}
\widetilde{\phi}={\cal L}\phi = \frac{1}{\sqrt {m^*}}\left[\frac{d}{dx} -
(\ln U)^{'}\right]\phi.
\end{equation}
It follows immediately from (17)
 that ${\cal L}{\cal U}=0$. In order to obtain the solution of the equation (4)
at the energy of transformation $\lambda $, we shall use the
second linear independent solution  to (\ref{H-eq}), namely
$\hat{\cal U}(x)={\cal U}(x)\int^{x}dx'|{\cal U}(x')|^{-2}$ where
the integration limits depend on the boundary conditions. In
particular, for regular solutions satisfying the boundary
conditions $\phi(x=0)=0,~~\phi'(x)|_{x=0}=1 $, the lower
integration limit is  $0$ and the upper one is $x$, while for the
Jost solutions the integration limits are $-\infty$ and $x$,
respectively. As a result we get
\begin{equation} \label{eta} \eta(x) ={\cal L}{\hat{\cal U}}(x)=\frac{1}{\sqrt {m^*(x)}}
\frac{1}{{\cal U}(x)}. \end{equation}
Once $\eta$ is found, one can get a second solution of (4)
 at the energy of transformation $\lambda$. By
using the Liouville's formula once more, one gets
\begin{equation}\label{heta} \hat\eta(x) =\eta(x)\int^{x}dx'|\eta^2|^{-1}
=\frac{1}{\sqrt {m^*(x)}{\cal U}(x)}\int^{x}dx'{\cal
U}(x')m^*(x'){\cal U}(x').
\end{equation}
Hence, the information about all solutions of the initial
equations (\ref{H-eq}) provides the knowledge of all solutions of
the transformed equations (\ref{H1-eq}). As in the case of
Schr\"odinger equation, the functions $\phi(x,{\cal E})$ and
$\widetilde{\phi}(x,{\cal E})$ correspond to  Hamiltonians ${\cal
H}$ and $\widetilde{\cal H}$, respectively,
 are related through the transformation operator ${\cal L}$ (see
 (17)). The difference is that in our case ${\cal L}$ includes the position-dependent
 mass. As a consequence, the new potential
 $\widetilde{V}$ and solutions $\widetilde{\phi}$ depend on the effective mass $m^*(x)$.
 The function
$\eta(x)$ defined by (\ref{eta}) at the energy of transformation
${\cal E}=\lambda$ cannot be normalized and this is the reason why
$\lambda$ does not belong to the discrete spectrum of
$\widetilde{\cal H}$. Therefore, Hamiltonians ${\cal H}$ and
$\widetilde{\cal H}$ are isospectral with one exception of the
bound state with the energy ${\cal E}=\lambda$, which is removed
from the initial spectrum of ${\cal H}$. Note that if the
transformation function ${\cal U}(x)$ corresponds to the ground
state, i.e., ${\cal U}(x)$ is nodeless, then the transformed
potential $\widetilde{V}(x)$ has no any new singularity, exept the
singularities due to $V(x)$ (of course, we asumme $m^*(x)\ne 0$ at
all $x$). However, if we apply this transformation to an arbitrary
state other than ground state, the transformed potential
$\widetilde{V}(x)$ might contain extra singularities, which are
not present in the initial potential $V(x)$ and hence, the
Hamiltonian $\widetilde{H}$ becomes physically meaningless. As we
shall see later, the difficulties with singularities can be
circumvented by means of second-order Darboux transformations. Now
we show how one can  construct a Hamiltonian with an additional
bounded state  with respect to the initial Hamiltonian by using
factorization of Hamiltonians and supersymmetry.

{\sf The  supersymmetry} is based on factorization properties of
Darboux transformation operators $\cal L$ and  $\cal L^+$.  The
definition of formally conjugate operators is
$D^\dag=(CQ)^\dag=Q^\dag C^\dag$ and
$(\frac{d}{dx})^\dag=-\frac{d}{dx}$. In our case, the scalar
product of functions is defined by not the standard way $(f,g)$
but with the weight of $m^*(x)$: $(f,g)_m=\int m^*(x)f(x)g(x)$. In
this case instead of operator $D^\dag$ it is necessary to consider
the operator $m^{*-1}D^\dag m^*$. Therefore the operator ${\cal
L}^{\dag}$ adjoint to ${\cal L}=\frac{1}{ \sqrt{m^*}}(\frac{d}{dx}
+ K)$ is determined as
\begin{eqnarray}\label{L-ad}
{\cal L}^{\dag}=\frac{1}{\sqrt {m^*}} \left(-\frac{d}{dx} -
\frac{m^{*'}}{2m^*} + K\right).
\end{eqnarray}
Now let us consider the superposition ${\cal L}^{\dag}{\cal L}$
and ${\cal L}{\cal L}^{\dag}$:
\begin{eqnarray}\label{Lp-L1}
&&{\cal L}^{\dag}{\cal L}=-\frac{1}{
m^*}\frac{d^2}{dx^2}+\frac{1}{m^*}(-K'+K^2),
\\
&&\label{L-Lp} {\cal L}{\cal L}^{\dag}=-\frac{1}{
m^*}\frac{d^2}{dx^2}+\frac{1}{m^*}(K'+K^2)-\frac{1}{2}\frac{m^{*''}}{m^{*2}}+
\frac{3}{4}\frac{m^{*'}m^{*'}}{m^{*3}}-\frac{m^{*'}}{m^{*2}}K.
\end{eqnarray}
Express the potential $V$ from equation (\ref{Eq-U}) in the
form $V={\cal U}''/(m^*{\cal U}) +\lambda $. Using $K'=-[{\cal
U}'/{\cal U}]'=-{\cal U}''/{\cal U}+({\cal U}'/{\cal U})^2$ we
represent $V$ as
\begin{eqnarray}\label{Vnew}
 V=\frac{1}{m^*}(-K'+K^2)+\lambda~.
 \end{eqnarray}
Substitution of (\ref{Vnew}) into (\ref{trV}) leads to the
following representation of the transformed potential:
\begin{eqnarray}\label{tr2}
\widetilde{V}=\frac{1}{m^*}(K'+K^2)+\frac{1}{\sqrt
{m^*}}\frac{d^2}{dx^2}\frac{1}{\sqrt {m^*}}-\frac{m^{*'}}{m^{*2}}K
+\lambda ~.
 \end{eqnarray}
Using (\ref{Vnew}) and (\ref{tr2}), after some
transformations the formulae (\ref{Lp-L1} and (\ref{L-Lp}) can be
rewritten as
\begin{eqnarray}\label{factor}
{\cal L}^{\dag}{\cal
L}=-\frac{1}{m^*}\frac{d^2}{dx^2}+V-\lambda={\cal
H}-\lambda;\\
\label{factor1} {\cal L}{\cal
L}^{\dag}=-\frac{1}{m^*}\frac{d^2}{dx^2}+\widetilde{V}-\lambda=\widetilde{\cal
H}-\lambda.
\end{eqnarray}
>From (\ref{factor1}) one can obtain
 the intertwining relation
\begin{equation}
\label{intert2} {\cal H}{\cal L}^{\dag}={\cal
L}^{\dag}\widetilde{\cal H}\,,
\end{equation}
which means that the operator ${\cal L}^{\dag}$ is also the
transformation operator and realizes the transformation of the
solutions of equation (\ref{H1-eq}) to solutions of (\ref{H-eq}),
$\phi\propto{\cal L}^{\dag}\widetilde{\phi}$. As one can see from
the comparison of the relations  (\ref{L1}) and (\ref{L-ad}), the
operator $ {\cal L}^{\dag}$ is not an inverse of $ {\cal L}$. One
can show that the operators ${\cal L}$ and ${\cal L}^{\dag}$ can
be expressed in terms of $\eta$, which are  solutions of transformed equations (\ref{H1-eq}) at the
energy $\lambda$  with the
potential $\widetilde{V}$ determined by (\ref{trV}). For this aim
let us express $K$ in terms of $\eta$, by means of (\ref{eta}).
$$ K=-\frac{{\cal U}'}{{\cal
U}}=\frac{m^{*'}}{2m^*}+\frac{\eta'}{\eta}.$$  Using this in
(\ref{L1}) and (\ref{L-ad}), we obtain
\begin{equation}
\label{L1-eta} {\cal L}= \frac{1}{\sqrt {m^*}}
\left(\frac{d}{dx}+\frac{m^{*'}}{2m^*} +\frac{\eta'}{\eta}\right),
~~~~~{\cal L}^{\dag}= \frac{1}{\sqrt {m^*}} \left(-\frac{d}{dx}
+\frac{\eta'}{\eta}\right)
\end{equation}
Evidently, the function $\eta$ is also a transformation function.
It is clear that ${\cal L}^{\dag}\eta=0$, i.e., $\eta$ belongs to
the kernel of the operator ${\cal L}^{\dag}$. As one can see from
(\ref{L1-eta}) and (\ref{heta}), the application of the operator
${\cal L}^{\dag}$ to the second linearly independent solution
$\hat \eta$ to equation (\ref{H1-eq}) gives back the solutions
${\cal U}$ of the initial problem at the energy of transformation.
Indeed, ${\cal L}^{\dag}\hat \eta=\frac{1}{\sqrt {m^*}}
\left(-\frac{d}{dx} +\frac{\eta'}{\eta}\right)
\eta(x)\int^{x}dx'|\eta^2|^{-1}={\cal U}$. Hence,  a one-to-one
correspondence between the spaces of solutions of equations
(\ref{H-eq}) and (\ref{H1-eq}) is established, and these are the  operators
${\cal L}$  and  ${\cal L}^{\dag}$, which produce the correspondence.

Note, one can interchange the role of the initial and final
equations. The function $\eta$ becomes transformation function for
the intertwining  operator  ${\cal L}^{\dag}$, which will make
the transformation in the opposite direction: from the potential
$\widetilde{V}$ to the potential $V$ and from the solutions of
(\ref{H1-eq})  to the solutions of (\ref{H-eq}). So, if within the
first procedure (\ref{L1})--(\ref{w-phi}) we constructed the
potential $\widetilde{V}$ with one  bounded state removed , now we
can construct the potential $V$ with an addional bounded state.

\subsection{Second-order and the chain of Darboux transformations}

Let us define the second-order Darboux transformation as a
sequence of two Darboux transformations performed in a row
 \begin{equation}
 \label{SL}
 {\cal L}={\cal L}_2{\cal L}_1,
\end{equation}
where ${\cal L}_1$  is actually  ${\cal L}$ defined in (14)
\begin{equation}
 \label{SL1}
 {\cal L}_1=\frac{1}{\sqrt {m^*}}\left(\frac{d}{dx}+ K_1\right),~~~
 K_1=-\frac{{\cal U}'_1}{\cal U}_1,
\end{equation}
whereas ${\cal L}_2$ is determined as follows:
\begin{equation}
 \label{SL2}
 {\cal L}_2=\frac{1}{\sqrt {m^*}}\left(\frac{d}{dx}+ K_2\right),~~~
 K_2=-\frac{\chi'_1}{\chi}_1,
\end{equation}
and $\chi_1\equiv\chi_1(x,\lambda_2)$ is obtained by means of the
first-order transformation, applied to the solution ${\cal U}_2$
of the equation (13)  or (\ref {H-eq}) with the eigenvalue $\lambda_2$
\begin{equation}
 \label{chi}
 \chi_1={\cal L}_1{\cal U}_2=\frac{1}{\sqrt {m^*}}\left(\frac{d}{dx}-
 \frac{{\cal U}'_1}{\cal U}_1\right){\cal U}_2.
\end{equation}
 It is clear that $\chi_1$ is the solution of equation (13) with the potential
 $V_1=V+2K_1'$, defined as in (\ref{trV}),
 and $\chi_1$ can be taken as a new transformation function for the Hamiltonian ${\cal H}_1$
to generate a new potential
\begin{equation}
 \label{SV2}
 V_2=V_1+\frac{1}{\sqrt {m^*(x)}}\left[\frac{d^2}{dx^2}\frac{1}{\sqrt {m^*(x)}}+
 2\frac{d}{dx}\left( \frac{1}{\sqrt {m^*(x)}}K_2\right)\right] \end{equation}
and corresponding solutions
\begin{equation}\label{Sphi2}
\phi_2={\cal L}_2\phi_1=\frac{1}{\sqrt {m^*}}
\left(\frac{d}{dx}+K_2\right)\phi_{1},~~\phi_{1}={\cal L}_1\phi.
\end{equation}
 Here the function $\phi_{1}$, denoted
earlier as $\tilde \phi$, is an eigenfunction of the Hamiltonian
${\cal H}_1$
\begin{equation}\phi_{1}= \frac{1}{\sqrt
{m^*}}\left[\frac{d}{dx} - (\ln U)^{'}_1\right]\phi.\end{equation}
In other words, the action of the second-order operator (\ref{SL})
on the solutions $\phi$ leads to the solutions of ${\cal H}_2$
\begin{equation}
\phi_2={\cal L}\phi={\cal L}_2{\cal L}_1\phi.
\end{equation}

Iterating this procedure $m$ times in regard to given operator
${\cal H}$, one arrives at the operator  ${\cal H}_m$, which
satisfies the intertwining relation
$${\cal L}{\cal H}={\cal H}_m{\cal L}.$$
In this way one gets
\begin{equation}\label{VM-pr}
 V_m=V_{m-1}+\frac{1}{\sqrt {m^*}}\left[\frac{d^2}{dx^2}\frac{1}{\sqrt {m^*}}+
 2\frac{d}{dx}\left( \frac{1}{\sqrt {m^*}}K_{m-1}\right)\right]
 ,
 \end{equation}
 \begin{equation}\label{phi-m}
 \phi_m={\cal L}\phi={\cal L}_m\phi_{m-1}={\cal L}_m{\cal L}_{m-1}...{\cal L}_1\phi,
\end{equation}
where ${\cal L}$ is the $m$-th order differential operator:
\begin{equation}\label{LM}
{\cal L}={\cal L}_m{\cal L}_{m-1}...{\cal L}_1,~~{\cal L}_m =\frac{1}{\sqrt {m^*}}\left(\frac{d}{dx}+K_m\right),
~~K_m=-\chi'_{m-1}\chi^{-1}_{m-1}.
\end{equation}
It should be noted, that the chain of $m$ first-order Darboux
transformations results in a chain of exactly solvable Hamiltonians
${\cal H}\to {\cal H}_1\to...\to{\cal H}_m$.

Consider now the 2-nd order transformation in  detail. Using the
explicit expression for $V_1$ which appears in the first-order
Darboux transformation, we get a formula for the potential $V_2$:
\begin{equation}\label{V2K}
V_2=V + \frac{2}{\sqrt {m^*}}\left(\frac{d^2}{dx^2}\frac{1}{\sqrt
{m^*}}\right)+ \frac{2}{\sqrt {m^*}}\frac{d}{dx}\left(
\frac{1}{\sqrt {m^*}}K\right),
\end{equation}
 where $K=K_1+K_2$.  Let us represent
$\chi_1$ as
\begin{equation} \label{chi1}
\chi_1(x)=\frac{1}{\sqrt {m^*(x)}}\frac{W_{1,2}(x)}{{\cal
U}_1(x)},
\end{equation}
 where $W_{1,2}(x)={\cal U}_1(x){\cal U}'_2(x)-{\cal U}'_1(x){\cal U}_2(x)$ is
 the Wronskian of the functions
  ${\cal U}_1(x)$ and ${\cal U}_2(x)$.    Plugging (\ref{chi1}) into
  the formula (\ref{SL2}) for $K_2$, after some transformations we obtain
  \begin{equation}\label{SK-2}
  K_2(x)= -\frac{d}{dx}\left[\ln \frac{W_{1,2}(x)}{\sqrt {m^*(x)}{\cal
  U}_1(x)}\right].
  \end{equation}
  After this $K=K_1+K_2$ can be represented as
 $$ K=-\frac{{\cal U}'_1}{{\cal U}_1}+\frac{m^{*'}}{2m^*}+\frac{{\cal U}'_1}{{\cal
 U}_1}-\frac{W'_{1,2}}{W_{1,2}} =\frac{m^{*'}}{2m^*}-\frac{W'_{1,2}}{W_{1,2}}.$$
With this taking into account, making in (\ref{V2K}) the next substitution:
$$\frac{1}{2}\frac{d}{dx}\frac{m'}{m^{*3/2}}=-\frac{d^2}{dx^2}\frac{1}{m^{*1/2}},$$
after some manipulations the new potential can be expressed as:
\begin{equation}\label{V2tr}
V_2(x)=V(x) -\frac{2}{\sqrt {m^*}}\frac{d}{dx}\left[\frac{1}{\sqrt
{m^*}}\frac{d}{dx}\ln W_{1,2}(x)\right].\end{equation} By using
(\ref{Sphi2}) find now the corresponding functions $\phi_2(x)$,
$\phi_{2}=\left(\frac{d}{dx}+K_2\right)\phi_{1} $. By analogy with
$\chi_1$ the function $\phi_1(x)$ can be written in terms of the
Wronskian $W_{1,{\cal E}}(x)={\cal U}_1(x)\phi'({\cal E},x)-{\cal
U}'_1(x)\phi({\cal E},x)$:
\begin{equation} \label{phi1}
\phi_1(x)=\frac{1}{\sqrt {m^*(x)}}\frac{W_{1,{\cal E}}(x)}{{\cal
U}_1(x)}.
\end{equation}
Let us now calculate the derivative of $\phi_1={\cal L}_1\phi$,
that is
$$ ({\cal L}_1\phi)'=
\frac{1}{\sqrt{m^*}{\cal U}'_1} + \frac{1}{\sqrt{m^*}}\phi''-
\frac{1}{\sqrt{m^*}}\frac{{\cal U}''_1}{{\cal U}_1}\phi.
$$
Making use of the last expression and the relation (\ref{SK-2})
for $K_2$, we obtain, after some simplification, the  formula
\begin{equation}\label{S-phi-2}
\phi_2(x)=\frac{1}{m^*(x)}\left(\phi''(x)-\frac{{\cal
U}''_1(x)\phi(x)}{{\cal U}_1(x)}\right) -\frac{d}{dx}\Bigl(\ln
W_{1,2}(x)\Bigr)\frac{W_{1,{\cal E}}(x)}{m^*(x){\cal U}_1(x)}.
\end{equation}
It is easily seen from (\ref{V2tr}) and (\ref{S-phi-2}) that due
to the 2-nd order Darboux transformation, the potential and
solutions obtained in this way are completely expressed in terms
of the known effective mass function $m^*(x)$ and the solutions
${\cal U}_1(x), {\cal U}_2(x), \phi({\cal E},x)$ to the initial
equation, with no use of the solutions to the intermediate one
with the potential $V_1(x)$.

Clearly,  for the next transformation step to be made, one should
take  a new transformation function $\chi_2$, that corresponds to
the potential $V_2$. It can be obtained by applying the operator
${\cal L}={\cal L}_2{\cal L}_1$ to the solutions ${\cal U}_3$
corresponding to the eigenvalue  ${\cal V}_3$:
$$\chi_2=\frac{1}{m^*(x)}\left({\cal U}''_3-
\frac{{\cal U}''_1}{{\cal U}_{1}}{\cal U}_3\right)
-\frac{d}{dx}\left(\ln
W_{1,2}(x)\right)\frac{W_{1,3}(x)}{m^*(x){\cal U}_1(x)}.$$ Then it
can be used to produce a new transformed operator ${\cal
L}_3=d/dr+K_3,~~K_3=-\chi'_2\chi^{-1}_{2}$ for generating  new
potential $V_3$ and solutions $\phi_3$ and so on, according to
(\ref{VM-pr})--(\ref{LM}).

\subsection{The integral form of Darboux transformations}

The transformed solutions (\ref{phi1}) and (\ref{S-phi-2}) can be
represented in the integral form.  Let us consider to this end the
generalized Schr\"odinger equation written down as
\begin{equation}
\label{G-eq} -\phi''(x)+ m^*(x)V(x)={\cal E}m^*(x)\phi(x).
\end{equation}
Multiplying both sides of the equation (\ref{G-eq}) for the function $\phi({\cal
E},x)$ by ${\cal U}_1(x)$
at the energy of transformation $\lambda_1$
and subtracting from the obtained expression  the equation similar to (\ref{G-eq}) but
written down for ${\cal U}_1(x)$ and
multiplied by  $\phi({\cal E},x)$, we arrive at
\begin{equation}
\label{W-1E} \frac{d}{dx}W_{1,{\cal E}}(x)=(\lambda_1-{\cal
E})m^*(x){\cal U}_1(x)\phi({\cal E},x).
\end{equation}
The last expression can be easily integrated:
\begin{eqnarray}
\label{wron-in} W_{1,{\cal E}}(x)= (\lambda_1-{\cal
E})\int_{a}^{x}m^*(x'){\cal U}_1(x')\phi(x')dx'+ C .
\end{eqnarray}
Inserting the last expression into the formula for $\phi_1$
(\ref{phi1}), we arrive at the integral form of the 1st order
transformed solutions:
 \begin{eqnarray}
\label{Phi1-int} \phi_1(x)= \frac{\left[C+(\lambda_1-{\cal
E})\int_{a}^{x}m^*(x')
 {\cal U}_1(x')\phi(x')dx'\right]}{m^*(x){\cal U}_{1}(x)} .
\end{eqnarray}
Here $C$ and $a$ are some arbitrary  constants. By analogy, applying this technique to
the equation
(\ref{H-eq}) for $\phi$ and ${\cal U}_{1}$,
using
(\ref{wron-in}) in (\ref{S-phi-2}),  we get the integral
form for the 2-nd order transformed solutions
\begin{eqnarray}
\label{phi2-int}\phi_2= (\lambda_1-{\cal E})
\phi(x)-\frac{d}{dx}\Bigl(\ln
W_{1,2}(x)\Bigr)\frac{\left(C+(\lambda_1-{\cal
E})\int_{a}^{x}m^*(x'){\cal U}_1(x')\phi(x')dx' \right)}{m^*(x)
{\cal U}_1(x)}.
\end{eqnarray}
Here the integration limits depend on the boundary conditions. In
particular, for regular solutions satisfying the boundary
conditions $\phi(x=0)=0, \phi'(x)|_{x=0}=1$, the lower integration
limit is $0$ and the upper one is $x$, while for the Jost
solutions the integration limits are $x$ and  $\infty$,
respectively. The constant $C$ is determined by the values of the
Wronskian at zero or at infinity, depending on the way the problem
is posed. Notice that the functions ${\cal U}$ and $\phi$ can be
chosen in such a way that the constant $C$ becomes zero.
Analogously to  (\ref{W-1E}), one has
$W'_{1,2}(x)/(\lambda_1-\lambda_2)=m^*(x){\cal U}_{1}(x){\cal
U}_{2}(x)$ and
\begin{eqnarray}
\label{W12}\frac{W'_{1,2}(x)}{W_{1,2}(x)}=\frac{m^*(x){\cal
U}_{1}(x){\cal U}_{2}(x)}{c_1+\int^{x}dx'm^*(x'){\cal
U}_{1}(x'){\cal U}_{2}(x')}.
\end{eqnarray} Using the last formula and assuming $C=0$, after some transformations one
can represent $\phi_2$  as follows:
\begin{eqnarray}
\label{phi2-int-n}\phi_2= (\lambda_1-{\cal E}) \phi(x)- \frac{
(\lambda_1-{\cal E}){\cal U}_2(x)\int_{a}^{x}m^*(x'){\cal
U}_1(x')\phi(x')dx' }{c_1 +\int^{x}dx'm^*(x'){\cal U}_{1}(x'){\cal
U}_{2}(x')}.
\end{eqnarray}

 Now let us consider the 2-nd order Darboux
transformation at $\lambda_1=\lambda_2\equiv \lambda$. Earlier
within the first-order procedure, we already obtained two linear
independent solutions (\ref{eta}) and (\ref{heta}) at
$\lambda_1=\lambda_2$. The second transformation can be made by
means of a linear combination of the solutions $\eta$ and
$\hat{\eta}$
\begin{eqnarray}\label{chi1-int}
\chi_1(x)=c_1\eta(x)+\widehat{\eta}(x)= \frac{1}{\sqrt
{m^*(x)}{\cal U}(x)}\left(c_1+ \int^{x}dx'{\cal
U}^2(x')m^*(x')\right).
\end{eqnarray}
In order to find the transformed potential and solutions, calculate
$K_2=-\chi'_1/\chi_1$
and $K=K_1+K_2$
$$K(x)=\frac{m^{*'}(x)}{2m^*(x)}-
\frac{ m^*(x){\cal U}^{2}_{1}(x)}{\left(c_1+ \int^{x}dx'{\cal
U}^2(x')m^*(x')\right)}. $$ Plugging the last expression into the formula (\ref{V2K})
which defines
the potential,  we arrive at
\begin{eqnarray}
\label{V-int}  V_2(x)=
V(x)-\frac{2}{\sqrt{m^*(x)}}\frac{d}{dx}\left(\frac{1}{\sqrt{m^*(x)}}\frac{{\cal
U}^2(x)m^*(x)}{(c_1+ \int^{x}dx'{\cal U}^2(x')m^*(x')}\right).
\end{eqnarray}
 The operator ${\cal L}_2$ (\ref{SL2}) with
$\chi_1$ defined by (\ref{chi1-int}), acts on the function
$\phi_1$ represented by its integral form (\ref{Phi1-int}) so that
it leads to
\begin{eqnarray}
\label{phi-int}\phi_2= (\lambda-{\cal E}) \phi(x)-\frac{ {\cal
U}(x)(\lambda-{\cal E})\int^{x}dx'{\cal U}(x')m^*(x')\phi(x')} {
c_1+\int^{x}dx'{\cal U}^2(x')m^*(x')}.
\end{eqnarray}
It is worth mentioning, that the formulae for the new potential
$V_2$ and the solution $\phi_2$ can be obtained directly  from the
relations (\ref{V2tr}) and (\ref{S-phi-2}), if one takes into
account that at $\lambda_1=\lambda_2\equiv \lambda$, the
expression (\ref{W12}) for $\frac {d}{dx}\ln W_{1,2}(x) $ should
be changed by
$$\frac{d}{dx}\ln P(x)=
\frac{m^*(x){\cal U}^2(x)}{ c_1+\int^{x}dx'{\cal U}^2(x')m^*(x')},
$$
with $ P(x)=c_1+\int^{x}dx'{\cal U}^2(x')m^*(x')$.

Without loss of generality one can take the linear combination of
the functions $\eta$ and $\hat{\eta}$ as
$\chi_1(x)=\eta(x)+C\widehat{\eta}(x)$, and change $(\lambda-{\cal
E}) \phi(x)\to  \phi(x) $ for simplification. Then formulae
(\ref{V-int}) and (\ref{phi-int}) can be  rewritten as
\begin{eqnarray}
\label{V-int-cor}  V_2(x)=
V(x)-\frac{2}{\sqrt{m^*(x)}}\frac{d}{dx}\left(\frac{1}{\sqrt{m^*(x)}}\frac{C{\cal
U}^2(x)m^*(x)}{(1+ C\int^{x}dx'{\cal U}^2(x')m^*(x')}\right).
\end{eqnarray}
\begin{eqnarray}
\label{phi-int-cor}\phi_2=  \phi(x)-\frac{ {\cal
U}(x)C\int^{x}dx'{\cal U}(x')m^*(x')\phi(x')}
{1+C\int_{x_o}^{x}dx'{\cal U}^2(x')m^*(x')}.
\end{eqnarray}
The constant  $C$  plays the role of a normalization constant or
the difference between the normalization constants of the bound
state $\lambda$ for the potentials $V_2(x)$ and $V(x)$,
respectively. Notice, the choice of arbitrary constants $x_o$ and
$C$ allows one to avoid the problems with zero-equal denominators,
or in other words, it means that one can make transformations on
an arbitrary bounded state and construct the potential without
singularities. Notice also, that $m^*(x)$ itself does not lead to
the singularities, because  the effective mass $m^*(x)\ne 0 $  and
assumed to be smooth and at least twice
 differentiable function with respect to space-variable.

The solution of the equation (\ref{H-eq}) with the
potential (\ref{V-int-cor}) at the energy of transformation
$\lambda $ can be achieved  by means of operator ${\cal L}_2$ acting on the solution
$\eta$ from (\ref{eta}), obtained within the first transformation
step
 $$\eta_2(x) ={\cal
L}_2\eta=\frac{1}{\sqrt
{m^*(x)}}\left(\frac{d}{dx}-\frac{\chi'(x)}{\chi(x)}\right)\frac{1}{\sqrt
{m^*(x)}}\frac{1}{{\cal U}(x)},$$ where $\chi'$  is assumed to be of the
form (\ref{chi1-int}). Finally we get
\begin{equation}\label{eta2}
\eta_2(x) =-\frac{C{\cal U}(x)}{1+C\int^{x}dx'm^*(x'){\cal
U}^2(x')}.
\end{equation}
One can rewrite the potential (\ref{V-int-cor}) and the solutions
(\ref{phi-int-cor}) in terms of $\eta_2(x)$ as
\begin{eqnarray}
\label{V-eta}
 & &
 V_2(x)=V(x)+\frac{2}{\sqrt{m^*(x)}}\frac{d}{dx}
[\sqrt{m^*(x)}
\eta_2(x){\cal U}(x)]\,,\\
\label{phi-eta} & & \phi_2(x)=\phi(x)+\eta_2(x) C\int^{x}dx'{\cal
U}(x')m^*(x')\phi(x').
\end{eqnarray}
The relations (\ref{V-int-cor}) -- (\ref{phi-eta}) are the results
of performing two subsequent transformations with the same energy.
Therefore, it allows one to construct the phase-equivalent
potentials. Indeed, if $C=N_2^{2}-N^2$ is the difference in
normalization constants of the bound state $\lambda$ for the
potentials $V_2(x)$ and $V(x)$ respectively, then the formulae
(\ref{V-int-cor}), (\ref{phi-int-cor}) and (\ref{eta2}) correspond
to phase-equivalent potentials whose scattering data  coincide
 and differ only by a
normalization factor. Note, the phase-equivalent potentials have a
different shape. They can be more deeper and
narrow or more shallow and wider and possess the same spectral
data, except for normalization constants.

  If we assume the
transformation function ${\cal U}(x)$ to be taken at the energy of
the bounded state, which we would like to add to the initial
spectrum, and $C=N^2$ is the corresponding normalization constant,
then the formulae (\ref{V-int-cor}), (\ref{phi-int-cor}) and
(\ref{eta2}) give the possibility to construct a potential with a
new bounded state $\lambda$ provided the other spectral
characteristics of the spectra produced by the potentials
$V_{2}(x)$ and $V(x)$, coincide. Notice, that the function ${\cal
U}(x)$, which is the solution of the initial equation with the
potential $V$, has to be taken at the energy of transformation
$\lambda$. To sum up, it can be said that by means of the
technique described above, it is possible to remove some bounded $\bigtriangledown$
states or to add  new ones and  to construct the phase-equivalent
potentials. The procedure can be repeated as many times as it is
needed to construct a new potential with a desirable spectrum.
Finally,  it should be noted that using the technique presented in
[12,13], it is not difficult to generalize these results to
include a case when the second order transformation is applied
simultaneously to $N$ bounded states. This, however, is beyond the
scope of the paper and will be a subject of another publication.

\section{Liouville transformation and QW potential reconstruction}

In previous sections we have reconstructed the QW potential in the
downright fashion using, in fact, the position-dependent potential
function. That is why we did not worry about the hermiticity of
the kinetic energy operator.

In this Section we propose a somewhat different approach to the
reconstruction of QWs potentials, which is based on the following
observations. First, in most of cases the number of bounded states
in QW is limited from above by some reasonable value of about 6-10
which, in its turn, is determined by the technological
limitations. Secondly, in most of  cases
 one can safely suppose that effective mass $m^*(x)$ is a smooth and at least twice
 differentiable, weakly position-dependent function. Therefore, we can use the
 {\it Liouville transformation} together with algebraic Darboux transformations in order
 to reconstruct the QW potentials with a predetermined spectrum.
This approach is applicable to the case of the straightforward reconstruction of QW
potential discussed previously, as well as to the case when the Hermitian kinetic energy
operator enters explicitly the generalized Schr\"odinger equation.

Let start from the equation of the form
\begin{equation}\label{Liou-1}
-\frac{d^2}{dq^2}\Phi(q)=m^*(q)\lambda\Phi(q), \end{equation}
where $q$ stands for space variable.
 Thus, one
can interpret this equation as a generalized Schr\"{o}dinger
equation with the potential $V\equiv 0$. Introduce now  new
variables and a new function as follows (this is one of the
special cases of Liouville transformation ):
\begin{equation}\label{Liou-2} m^*=p^2,\;
\sqrt{m^*}=\frac{dx}{dq},\; \Phi(q)=p^{-1/2}\Psi(x).
\end{equation} Then, taking into account (\ref{Liou-2}), the equation (\ref{Liou-1})
can be reduced to the following:
\begin{displaymath} -p^{3/2}\left[ \Psi^{''}(x)+
\frac{1}{2}\left\{\frac{1}{2}\left(\frac{p^{'}_{x}}{p}\right)^2-\frac{p^{''}_{x}}{p}\right\}
\Psi(x)\right]=p^2\lambda p^{-1/2}\Psi(x),\end{displaymath}
 which can be written down as a standard Schr\"{o}dinger equation
\begin{displaymath}-\Psi^{''}(x) + V(x)\Psi(x)=\lambda\Psi(x),
\end{displaymath} where the potential $V(x)$ is of the form:
 \begin{displaymath}
V(x)=\frac{1}{2}\left[\frac{p^{''}_{x}}{p}-\frac{1}{2}
\left(\frac{p^{'}_x}{p}\right)^2 \right].
\end{displaymath} Since
\begin{displaymath} \frac{p^{''}}{p}=\frac{d}{dx}\left(\frac{p^{'}}{p}\right)+
 \left(\frac{p^{'}}{p}\right)^2,
\end{displaymath}
the potential $V(x)$ can also be represented as
\begin{equation}
 \label{Liou-3}
 V(x)=\frac{1}{2}\left[\frac{d^2}{dx^2}\ln p +
\frac{1}{2}\left( \frac {d}{dx}\ln p\right)^2\right],
\end{equation} or returning back to the variable $m^*$,
\begin{displaymath}
V(x)=\frac{1}{4}\left[\frac{m^{*''}}{m^*}-
 \frac{3}{4}\left(\frac{m^{*'}}
{m^*}\right)^2\right].\end{displaymath}
 Now assuming the solution
of the eigenvalue problem to the Schr\"{o}dinger equation with
this potential is known, one can add or subtract one by one as
many bounded states as it is needed, using the algebraic Darboux
transformations described in the previous Section.

As for the effective mass theory with the Hermitian kinetic energy
operator is considered, it seems that a consensus among the
specialists has been already achieved, since in the majority of
corresponding studies  the following form of the Hamiltonian is
used (see, for example, [10]):
\begin{displaymath}
H= \left[\hat{P}\frac{1}{m^*(q)}\hat {P}\right] + V(q),\end{displaymath} where
$\hat{P}$ is the momentum operator. Then, in order to use the Liouville
transformation approach, one should start with the equation
\begin{displaymath}
-\frac{d}{dq}\left\{\frac{1}{m^*(q)}\frac{d}{dq}\Phi\right\}=\lambda
\Phi. \end {displaymath}
 Making the next substitutions
 \begin{displaymath} p=\frac{1}{\sqrt{m^*(q)}}=\frac{dq}{dx},~~~ \Phi=p^{-1/2}\Psi \end{displaymath}
 after some manipulations similar to that which were previously done, one get the
 standard Schr\"{o}dinger equation
 \begin{displaymath}
 -\Psi^{''}(x) + V(x)\Psi(x)=\lambda\Psi(x),
\end{displaymath} with the potential defined as in (\ref{Liou-3}) but with
$p(x)=m^{*-1}(x)$. Finally we obtain
\begin{displaymath}
V(x)=\frac{1}{4}\left[-\frac{m^{*''}}{m^*}+
 \frac{5}{4}\left(\frac{m^{*'}}
{m^*}\right)^2\right].\end{displaymath}

\section{Conclusion}

 The basic elements of contemporary micro- and nanoelectronics are the low-dimensional
 structures which are the structures composed of QWs, quantum wires and  quantum dots
 and produced by means of various techniques including most impressive one, molecular
  beam epitaxy. The entirety of such methods and techniques are sometimes termed as
 {\it Quantum Engineering} or {\it Quantum Technology}.
 One of the most important issues of quantum engineering is the construction of
 multi-quantum well structures possessing  desirable properties. This problem appears
 in different contexts, ranging from the  construction of multi-level computer logic
 to photovoltaics of third generation [18,19]. From the theorist's point of view, the
  problem can be formulated as follows: assume one requires a definite spectrum
  of QW, because it is determined by some specific needs and  circumstances. Can one
  reconstruct the QW potential which supports this very spectrum? In this paper
  we answer this question in affirmative and outline the possible
 strategy of the QW potential reconstruction, if the spectrum of QW is
predetermined.

  The proposed approach is based on the  combination of various
techniques such as {\it Inverse Scattering Problem Method}, {\it Darboux}
and {\it Liouville} transformation.
  Bearing in mind that the effective masses of charge carriers in the subsequent
   layers of different materials which make QW, are different, we match the
intertwining operator  technique, in order to take into account
the position-dependent mass in Eq. (\ref{H-eq}). The first- and
second-order of Darboux transformations, as well as the chain of
Darboux transformations are considered, and interrelation between
the differential and integral transformations is established. The
developed approach allows one to construct phase-equivalent
potentials and  to add (or if necessary, to remove) some states to
(or from) the spectrum supported by the initial potential, whose
form can be established for instance, by means of ISP-method.

 Another possible way to take into account the position-dependent masses of charge
  carriers is to use the  Liouville transformation. It also allows one to incorporate
   and treat on the same
footing the proper form of Hermitian kinetic energy operator. This
operator  appears in the context of applicability to
heterostructures the effective mass approximation, the subject of
today's research activities.

In the paper we only formulate the basic concepts and develop the
necessary mathematical tools. More thorough numerical examinations
of specific cases are required and it will be done elsewhere.

\section*{Acknowledgments}

One of the Authors (A.S) is acknowledged to the Russian Federation
Foundation for Basic Research for the partial support (grant
06-01-00228).

\end{document}